\documentclass[useAMS,usenatbib,letterpaper]{mn2e}
\usepackage[T1]{fontenc}

\title[Two low-mass planets in the presence of a gas giant]
{How the presence of a gas giant affects the formation of  
mean-motion resonances between two low-mass planets in a
locally isothermal gaseous disc}
\author[E. Podlewska-Gaca and E. Szuszkiewicz]
{E. Podlewska-Gaca$$\thanks{E-mail:
edytap@univ.szczecin.pl (EP)} and E. Szuszkiewicz$$\thanks{E-mail: 
szusz@univ.szczecin.pl (ES)} \\
Institute of Physics and CASA*, University of Szczecin, ul. Wielkopolska 15,
70-451 Szczecin, Poland}
\begin{document}

\date{}

\pagerange{\pageref{firstpage}--\pageref{lastpage}} \pubyear{}

\maketitle

\label{firstpage}

\begin{abstract}
In this paper we investigate the possibility of a migration-induced
resonance locking in systems containing three planets, namely an Earth 
analog (1$M_{\oplus}$), a super-Earth (4$M_{\oplus}$) and a gas giant 
(one Jupiter mass). 
The planets have been listed in order of increasing
orbital periods.
All three bodies are embedded in a locally isothermal gaseous disc and 
orbit around a solar mass star. We are interested in answering the 
following questions: 
Will the low-mass planets form the same resonant structures with each 
other in the vicinity of the gas giant as in the case when the gas giant 
is absent? 
More in general, how will the presence of the gas giant affect the 
evolution of the two low-mass planets? 
When there is no gas giant in the system, it has been already shown that 
if the two low-mass  planets undergo a convergent differential migration, 
they will capture each other in a mean-motion resonance. 
For the choices of disc parameters and planet masses made in this paper, 
the formation of the 5:4 resonance in the absence of the Jupiter
has been observed in a previous 
investigation and confirmed here.
In this work we add a gas giant on the most 
external orbit of the system in such a way that its differential migration  
is convergent with the low-mass planets. We show that the result of this 
set-up is the speeding up of the migration of the super-Earth and, after 
that, all three planets become locked in a triple mean-motion resonance. 
However, this resonance is not maintained due to the low-mass planet 
eccentricity excitation, a fact that leads to close encounters between 
planets and eventually to the ejection from the internal orbits of one or 
both low-mass planets. 
We have observed that the ejected  low-mass planets can leave the system,
fall into a star or become the external planet relative to the gas giant.
In our simulations the latter situation has been observed for the 
super-Earth.  It follows from the results presented here that the presence 
of a Jupiter-like planet might have a strong influence on the architecture 
of planetary systems. Moreover, the planet ejections due to the gas giant 
action may lead to the formation of a population of low-mass freely 
floating planets. 
\end{abstract}

\section{Introduction}

The rapid progress in  instrumentation, observational techniques
and strategies has provided an enormous amount of data, which has
been used for intensive planet searches. The rate of planetary system
discoveries is impressive: After 20 years of observations,
 we know already over 1000 
confirmed planets and a few thousand of planet 
candidates \citep{borucki,batalha}. An even more compelling
motivation for the
research presented here is 
the growing evidence that multi-planet systems are the rule
rather than the exception. The number of such systems at the time of writing 
this paper has reached over 170. In some of them the planetary orbits remain
in mean-motion commensurabilities \citep[e.g.][]{lissauerb,steffen12a,
fabrycky12a,steffen12b}. For example, the authors of the recent
paper \citep{steffen12b} have reported
that in all thirteen systems confirmed by their investigations
there are planets near one of the first order mean-motion resonances. 
More precisely, they have found pairs
of planets near the 2:1, 3:2, 4:3, 5:4 and 6:5 commensurabilities. 
There is a strong indication that in some of those systems there could be
multiple pairs of planets near first order mean-motion resonances
which might form triple resonances. This seems to be the case of
Kepler-53 (4:2:1) or Kepler-60 (20:15:12).
The importance of 3-body resonances has been recently
discussed also by \cite{miga}
in the context of the dynamics of Kepler-11.
All the planets involved in the mean-motion commensurabilities
mentioned above are 
super-Earths with masses in the range of 3 - 10 $M_{\oplus}$.
This suggests that 
complex multiple 
resonant structures might form
in systems with low-mass planets only.
Systems of this kind are probably not rare, because
preliminary 
results of the Kepler mission seem to confirm
the expectations that low-mass planets are numerous \citep{ida}. 
The case of two resonant low-mass planets migrating in a gaseous
protoplanetary  
disc has been studied in details by \cite{papszusz, papszusz10}. 
Their main finding 
is that, once convergent migration takes place, it is very likely that
the planets end up in one of the mean-motion commensurabilities, whose type
is critically dependent on the migration rate. In the 
most commonly realized resonances the
values of $p$ appearing in the period ratio  $(p+1):p$ 
are greater when the planet masses are disparate than when they are equal.
This is a 
straightforward consequence 
of the difference in the migration rates. In fact, the differential
migration of equal  
mass planets in the Earth mass range is slower than that of disparate mass 
planets.  
An
analytic criterion for the occurrence and maintenance of resonances
between low-mass planets has been provided in Papaloizou and
Szuszkiewicz (2010).

However, it is licit to expect that most systems will contain
not only low-mass planets, but
also more massive planets like gas giants.
For this reason, the aim of this paper is to investigate the behaviour 
of pairs of planets
with masses in the Earth mass range, similar to those considered in 
\citet{papszusz}, when there is in the system
an additional, more massive planet. In other words,
we ask ourselves how the presence of a giant planet can influence the
dynamics of a system containing two low-mass planetary companions.
The continuous flow of observational information 
about the dynamics and structure of  such systems 
offers a strong motivation
for our investigations on the characteristics of the planetary 
configurations
formed during the early stages of the evolution
with particular attention given to resonant 
configurations.  

The situation is more complex if we have more than two planets, 
as it
was shown for example by \citet{marzari} or \cite{moeckel} for a three giant 
planet system.  
In the case of a system with three planets we can
observe indeed a variety of dynamical behaviours, such as planet merging,
orbital exchange or scattering of the planets. 
We study first the evolution of a system containing only
two low-mass  
planets embedded in a
gaseous protoplanetary disc. Then, we add a Jupiter-like gas giant 
on the external orbit relatively to both low-mass planets and compare 
the evolution of the low-mass planets before and after this addition.
We choose to put a gas giant on the external orbit in order to
create an  environment favourable for the formation of the commensurabilities
between the super-Earth and the gas giant \citep{paperI}. 
It is  very difficult  
to get a resonance capture in the opposite configuration
 \citep{paperII,paperIV}.

We expect that the evolution of the system might be
strongly dependent on the 
migration rate of the giant planet, so we perform our
simulations in 
discs with different viscosities, changing in this way the migration
speed of the 
gas giant. As it is shown by the simulations, it turns out
 that also the migration 
of low-mass planets can be
 affected by the viscosity of the disc.
A similar effect has been already observed by \cite{masset}. They have
explained it
by non-linearities of the flow around
a low-mass planet that can cause an additional torque from the co-orbital
region which might slow down its migration rate.

We find 
that in the locally isothermal discs the following sequence of events
occurs during
the 
evolution of 
a three-planet system consisting initially of two low-mass planets
(an Earth analog and a super-Earth with the mass of four Earth masses) 
and a gas giant (with the mass of Jupiter): The system evolves first
into an unstable triple  resonance, followed by the ejection of one
or both low-mass planets. Therefore, the two most frequent outcomes of
the system  
evolution are that
the gas giant remains the most internal planet or
one of the low-mass planets stays in  resonant configuration with the
gas giant.  
If similar conditions are common among the observed young planetary systems,
the planets ejected during the evolution of such systems might form 
a population of planets freely floating in the space. 

The paper is organized as follows. In Section 2 we describe the
methodology
followed in the numerical
simulations, while in Section 3 we present the behaviour of a system
with two 
low-mass planets, showing how the viscosity affects their evolution. 
The results of the simulations with a gas giant with two low-mass companions
are discussed in Section 4 and partly also in Section 5, where our
conclusions are also drawn.

\section{Description of the numerical simulations}
\label{numerics}

We have performed full 2D numerical simulations of planets embedded in a
gaseous protoplanetary disc using the hydrodynamical code
NIRVANA. For details on the numerical scheme employed in the code see 
\citet{nelson2000}. 
We use polar coordinates ($r, \phi$) with the origin located at the
position of the central star.

In our simulations the mass unit is the mass of the central star. The initial
semi-major axis of the super-Earth defines the unit of length. The unit of time
is the orbital period of the initial orbit of the super-Earth divided by
2$\pi$. In these units, the gravitational constant $G=1$.
With the above settings,
our computational domain extends from 0.33 to 5  and it
is divided uniformly into 489 and 512 grid cells in the radial and azimuthal directions
respectively. The radial resolution of the grid is therefore 
$\delta r = 9.6 \times 10^{-3}  = 0.192 H/r$, where $H$ is the semi-thickness 
of the
disc. This means that, before the low-mass planet motion becomes
dominated by the presence of the gas giant in the disc, i.e. when 
$r \ge 0.78$, $H$ 
is resolved by no less than 4 grid cells at the planet
locations. 
Instead, the width of the horseshoe region is covered just by one grid
cell, which indicates that the corotation torque is poorly resolved.
However, this should not affect our results, because we expect the corotation
torque to saturate under the conditions adopted in this study (see Section 
\ref{conclusions}).
The resolution of our grid is a compromise between making the size of the  
disc appropriate for  modelling our three-planet system and being able
to follow the evolution of planets for a sufficiently long period of time.  
In order to strengthen the validity of our results,
we have verified that the final outcome of our simulations does not change
if we double the resolution in each direction 
(978 and 1024 grid 
cells in the radial and azimuthal directions respectively).  
We choose open radial boundary conditions, so that
the material in the
disc can outflow through the boundaries of the computational domain
according to the viscous evolution of the disc.
The disc has constant aspect ratio $h=H/r=0.05$ and uniformly
distributed surface density $\Sigma_0=6\times10^{-4}$.
This value of $\Sigma_0$
corresponds to the minimum mass solar nebula (MMSN) around 5.2 AU if one
takes the length unit equal to 5.2 AU and the mass unit equal to the 
solar mass. 
We use the locally isothermal equation of state and 
do not take into account the disc self-gravity. Those assumptions have
been made in the previous paper \citep{papszusz}, which is the starting 
point for
the present study. This is why we have not employed here a  more realistic
equation of state and the disc self-gravity has been neglected.
The gravitational potential is softened with softening parameter
$\varepsilon=0.8H$ and we do not exclude any material within the Hill sphere.
\cite{crida2009} argue that the material inside the Hill sphere should be 
very carefully 
taken into account in numerical simulations when the self-gravity of the gas 
is neglected. 
How this should be realized in a proper way is still a subject of intensive
research.
However, to take fully into account self-gravity  
is not essential for the purpose of this study. We have 
rather concentrated on the
influence of the relative migration rate which has been tuned by 
the choice of different values of the viscosity.  
Finally, to close the list of assumptions, the planets do not
accrete matter from the disc. In all simulations in which there
is a gas giant in the system, all planets are kept on fixed
orbits until the 
giant planet opens a deep gap in the disc and only after that time they
are allowed to migrate. 

In the simulations presented here 
the central star has a Solar mass, so the masses of the planets are
respectively $m_E=1 M_{\oplus}$, $m_{SE}=4 M_{\oplus}$ and $m_{J}=1 M_{J}$,
where $M_{\oplus}$ is the Earth mass and $M_{J}$ is the Jupiter mass.
The planet with  mass $m_E=1
M_{\oplus}$ (called hereafter  ``the Earth analog'' or 
``the Earth-like planet'') is located initially
on the innermost circular  
orbit at the distance
$r_E=0.83$. 
Another
planet with the mass $m_{SE}=4 M_{\oplus}$ (which will be called "the
super-Earth") 
 is located initially at $r_{SE}=1.0$, also on the circular orbit. 
This 
configuration of the low-mass planets
is similar to the one described in \citet{papszusz}.
With respect to \citet{papszusz},
in the present work the kinematic viscosity $\nu$ is introduced
and the effects of the addition of a gas giant with the mass of Jupiter
in the system are
investigated. 
We have initially
placed the gas giant
at $r_J=1.62$, which enables the formation of any of the first 
order
mean-motion resonances. Then, we have moved its location to
$r_J=1.35$ and  $r_J=1.28$.
The kinematic viscosity
is changed
from  simulation to simulation and can take the values
$\nu=0, 2 \cdot 10^{-6}, 6 \cdot 10^{-6}$ and $10^{-5}$
expressed in our units.
The motion of the planets is calculated using a standard leapfrog integrator.

\section{The evolution of a system with two low-mass planets}
\begin{figure}
\centering
\vskip 5.0cm
\includegraphics{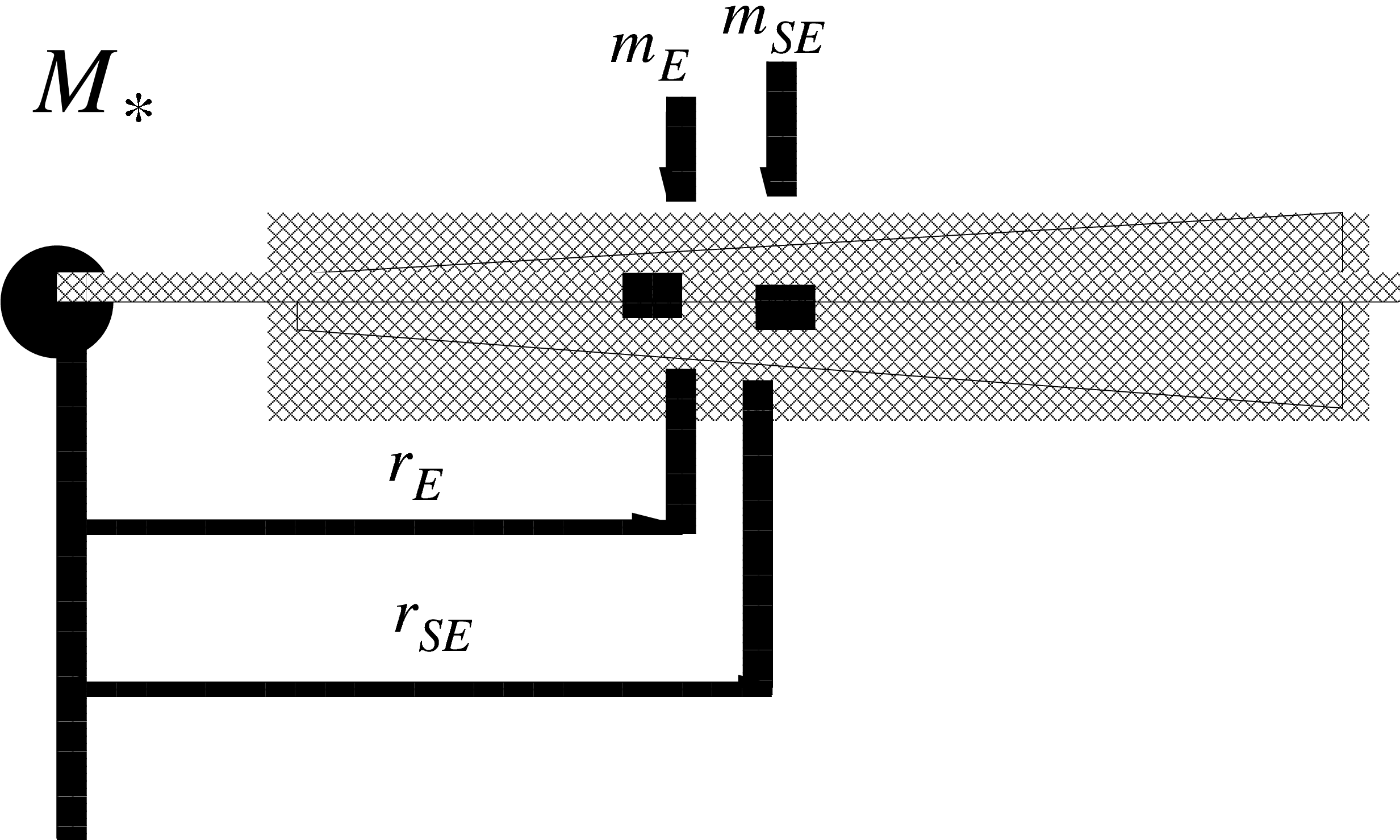}
\caption{\label{fig00}{The initial set up for numerical simulations
in the case of two low-mass planets embedded in a protoplanetary disc.
}}
\end{figure}

\begin{figure*}
\begin{minipage}{180mm}
\centering
\vskip 5.5cm
\includegraphics{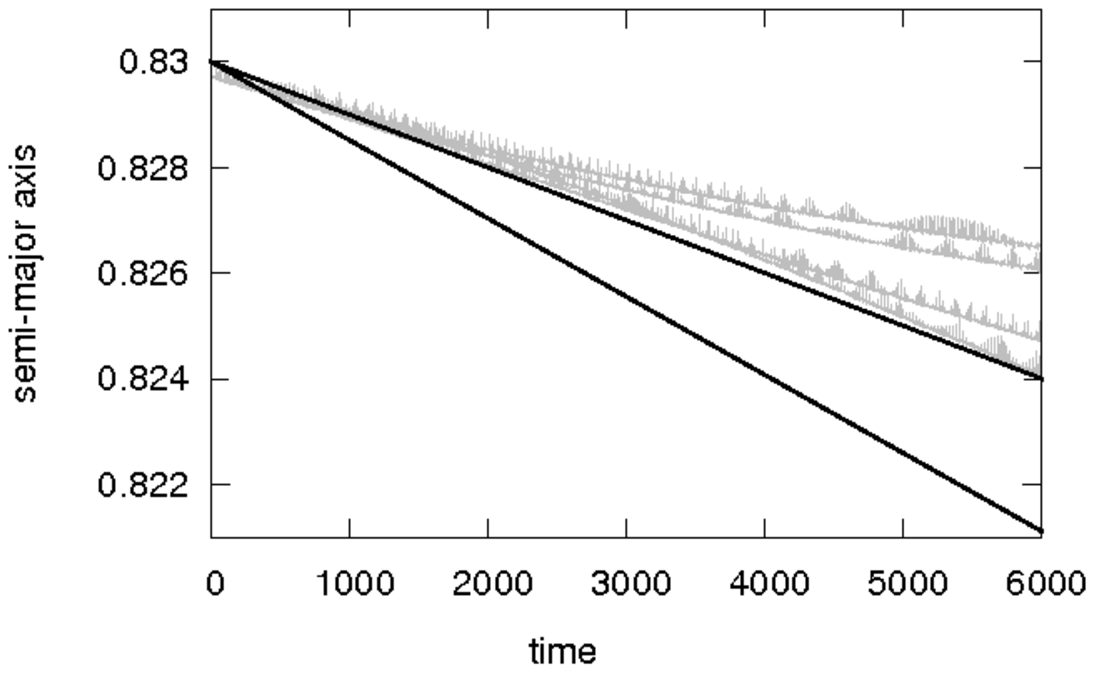}
\includegraphics{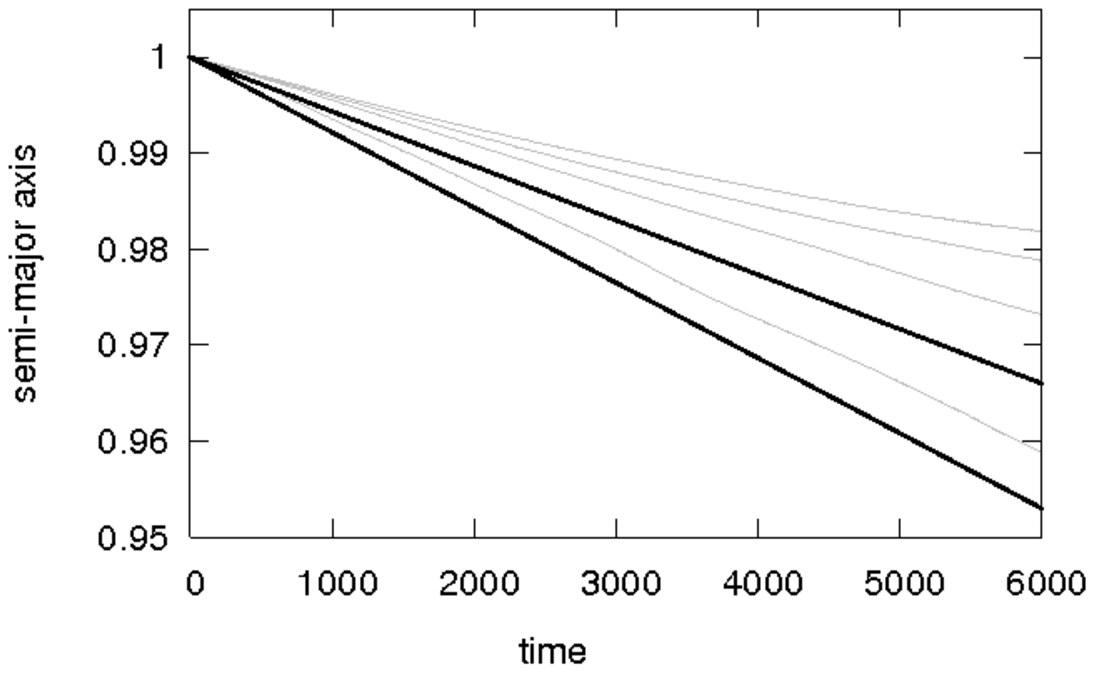}
\end{minipage}
\caption{\label{fig1}{The migration rates of planets
with the masses $1 M_{\oplus}$ ({\it left panel}) and $4 M_{\oplus}$ 
({\it right panel}) 
in
the discs with four different kinematic viscosities. The gray curves denote the
 results
 for $\nu=10^{-5}, 6 \cdot 10^{-6}, 2 \cdot
 10^{-6}$ and $0$ from top to bottom respectively. The upper and lower black
 lines denote the migration rates calculated according to \citet{Paa2010a}
(equation (49) in their paper)  and 
 \citet{tanaka02} (equation (70) in their paper) respectively.
}}
\end{figure*}

First of all, we investigate the behaviour of a system containing 
two low-mass planets embedded in a gaseous protoplanetary disc.
This will be a useful reference for comparison with the full
three-planet system that is the main goal of this study.
The initial set up for our simulations is illustrated in Fig. \ref{fig00}.
Before adding a gas giant to the system, we check the dependence of the 
planet migration rate 
on the value of the kinematic
viscosity of the disc. The values of $\nu$ range  
from $0$ to $10^{-5}$, so that we can control the effects
of the viscosity on the planet migration. This will allow us
to perform in the next 
Section
a series of simulations
with different migration speeds of the gas giant while retaining
 full control of 
possible changes caused by the viscosity in the low-mass planet migration.
The migration rates of the two low-mass planets in discs with
four different  
values of viscosity
are shown in Fig. \ref{fig1} separately for the Earth analog and the
super-Earth before the two planets start to interact strongly with each other.
\begin{figure*}
\begin{minipage}{180mm}
\centering
\vskip 5.5cm
\includegraphics{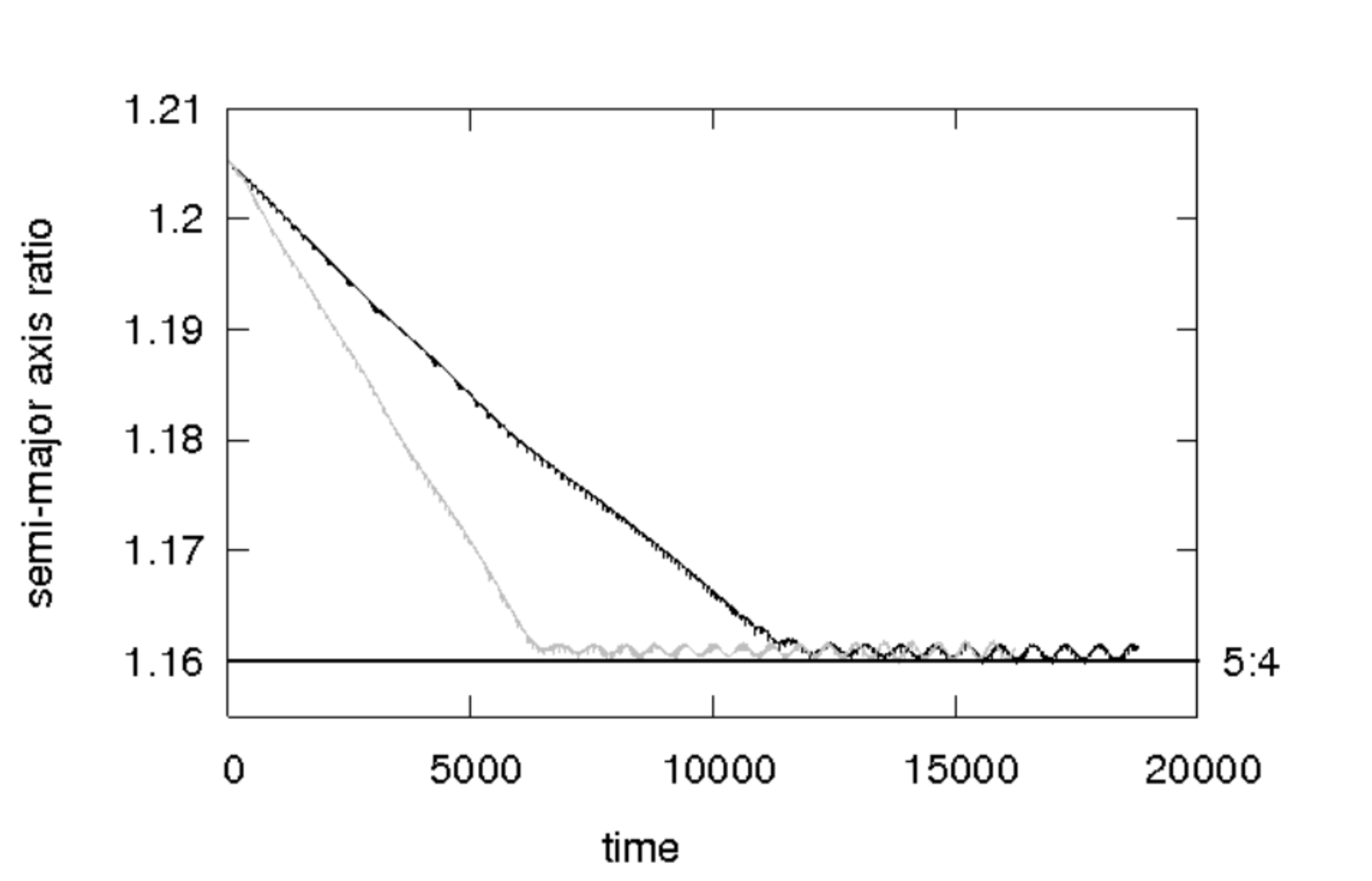}
\includegraphics{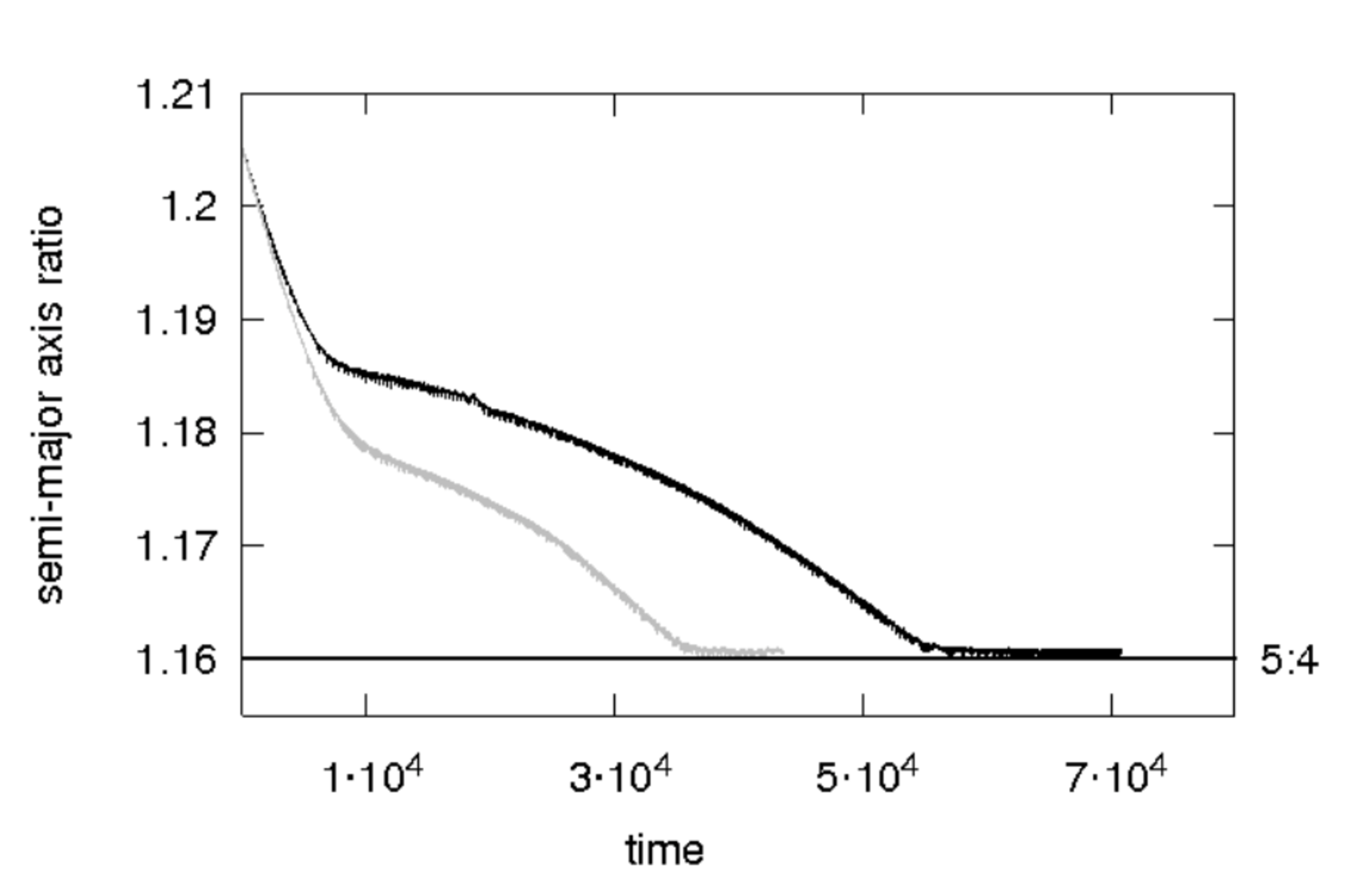}
\end{minipage}
\caption{\label{fig2}{The left panel present the semi-major axis ratios
of the planets embedded
 in a disc with the kinematic viscosities
 $0$ (gray colour) and $2 \cdot 10^{-6}$ (black colour).
The right panel shows the semi-major axis ratios of the planets embedded in 
the disc with the viscosities $6 \cdot
10^{-6}$ (gray colour) and $\nu=10^{-5}$ (black colour) where the differential
 migration is so slow that planets attain the 5:4 resonance after a longer
evolution time.
}}
\end{figure*}

The left panel presents the migration rate of the inner planet with mass
of $1 M_{\oplus}$. 
We can see that for 
lower viscosity the migration is faster. In the right panel of Fig.
\ref{fig1}
we plot the migration rates of the outer planet with the mass of 
$4 M_{\oplus}$. 
For both planets the migration rate changes with the viscosity in a similar
quantitative way, but 
the migration of the super-Earth is affected by the viscosity
much stronger than that of the less massive planet. The lowest black line
appearing in the left panel and that in the right panel
denote the migration rates of the planets
 with masses $1 M_{\oplus}$
and $4 M_{\oplus}$ respectively
calculated according to well known
3D prescription of the type I migration given by equation (70) 
in \cite{tanaka02}. 
However,
recent developments in the calculations of the torque acting on low-mass
planets
(\cite{PaarMel2006, BarMas2008, KleyCrida2008, Paa2010a, Paa2010b}) 
have revealed 
that the speed and even the direction of the migration
can be changed if we take into account a more refined treatment of
the disc structure. 
According to \cite{Paa2010a}, for the locally isothermal approximation
which
we use here, we expect to have a slower migration rate than those 
described
in \cite{tanaka02}. This is illustrated  
in  Fig. \ref{fig1}, where the upper black
lines  denote the migration rates obtained for
locally isothermal discs according  to the formula (49) of \cite{Paa2010a}.
Our simulations fits better to the upper black 
lines obtained after \cite{Paa2010a} than to the lower ones by \cite{tanaka02}.
Despite  the differences in the migration rates due to the disc viscosity,
in all our simulations the planets
are captured eventually in the 5:4 resonance.
To illustrate this, in Fig. \ref{fig2} we have plotted the 
evolution of the semi-major axis 
ratios of the two low-mass
planets in discs with four different values of the viscosity.
The results of our simulations are presented in  two separate panels, because 
the resonant capture
occurs at different timescales. The resonance capture in the discs
with 
the 
two 
lower values
of the viscosity ($\nu=0$ and $\nu=2\cdot 10^{-6}$) took place  after
6000 and 11000 time units respectively, see left panel. In
the discs with the higher values of the viscosity  ($\nu=6\cdot 10^{-6}$
and $\nu=10^{-5}$) it occurred
after 40000 and 50000
time units respectively, see right panel. 
 Note that the same resonance (5:4) obtained here has been
reported by  
\citet{papszusz} for planets
embedded in 
an inviscid disc with the same surface density as in our simulations.
In Fig. \ref{fig2} (right panel) one can observe another interesting
phenomenon, 
namely the passage through one of the higher
order resonances (most likely 9:7) in the case of viscosity
$\nu=10^{-5}$ (black curve) at around $2 \cdot 10^4$ time units. 
 
\begin{figure}
\centering
\vskip 5.5cm
\includegraphics{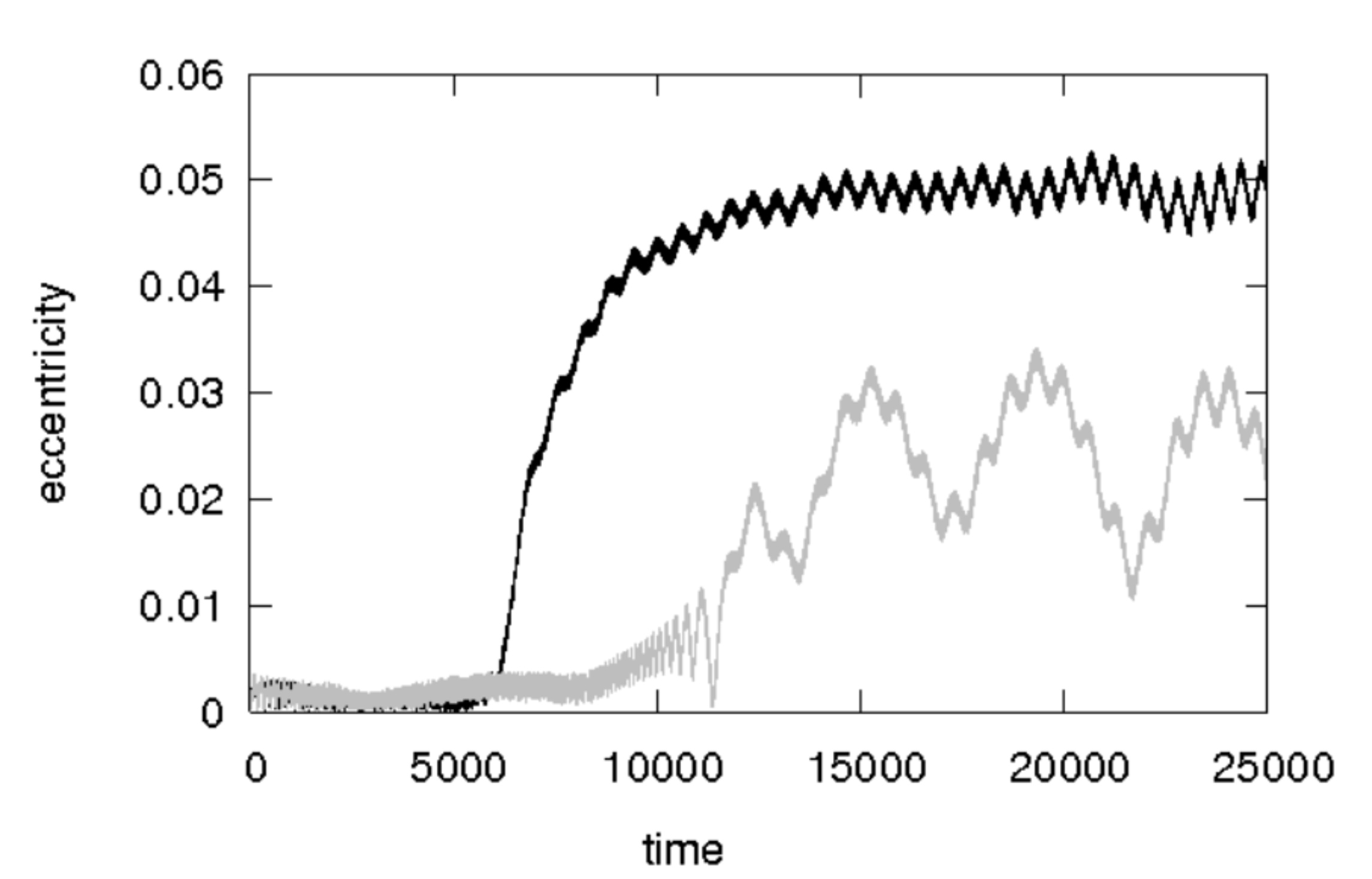}
\caption{\label{ecc}{The evolution of the eccentricities of the Earth-like
planet in the disc with the viscosity $2\cdot 10^{-6}$ (gray colour) and 
$0$ 
(black colour), respectively.
}}
\end{figure}
As soon as the capture in the mean-motion resonance occurs,
the eccentricities of the planets
increase. Assuming that the eccentricity of the super-Earth $e_{SE}$ is
negligible, which is true for our simulations ($e_{SE}\approx 0.007$),
\citet{papszusz} gave an analytic prescription for the final value
of the eccentricity of the inner planet $e_E$  in the inviscid disc
\begin{eqnarray}
{e_E}^2 & = & 
\left(\frac{m_{SE}}{m_E}\left(\frac{a_{SE}}{a_E}\right)^{1/2} - 
1\right)  \times \nonumber \\ 
& & \frac{m_E}{0.578(p+1)(m_E+m_{SE})}\left(\frac{H}{r}\right)^2,
\label{mimwzor}
\end{eqnarray}
where $p$ is an integer. 
 According to this estimate,
the eccentricity should reach 0.048 in the 5:4 resonance ($p$=4), which is
in very 
good 
agreement with
the eccentricity of the Earth-like planet in an inviscid disc 
shown in black colour in Fig. \ref{ecc}. 
As mentioned previously, if the viscosity in the disc is higher, the
migration 
of low-mass planets is slower. 
Thus, also the
eccentricity should reach lower values in  discs
with higher viscosity, as it is indeed observed in our
simulations. For example, in Fig. \ref{ecc} it is shown (gray colour line)
the evolution of the eccentricity of the Earth-like planet in the disc with
 viscosity $2\cdot 10^{-6}$. A similar correlation between the migration
rate and the eccentricity evolution was observed in \citet{crida} for 
systems of two giant planets.

\section{The evolution of two low-mass planets in the presence of a gas
giant}
\label{gasgiant}
\begin{figure}
\centering
\vskip 5.5cm
\includegraphics{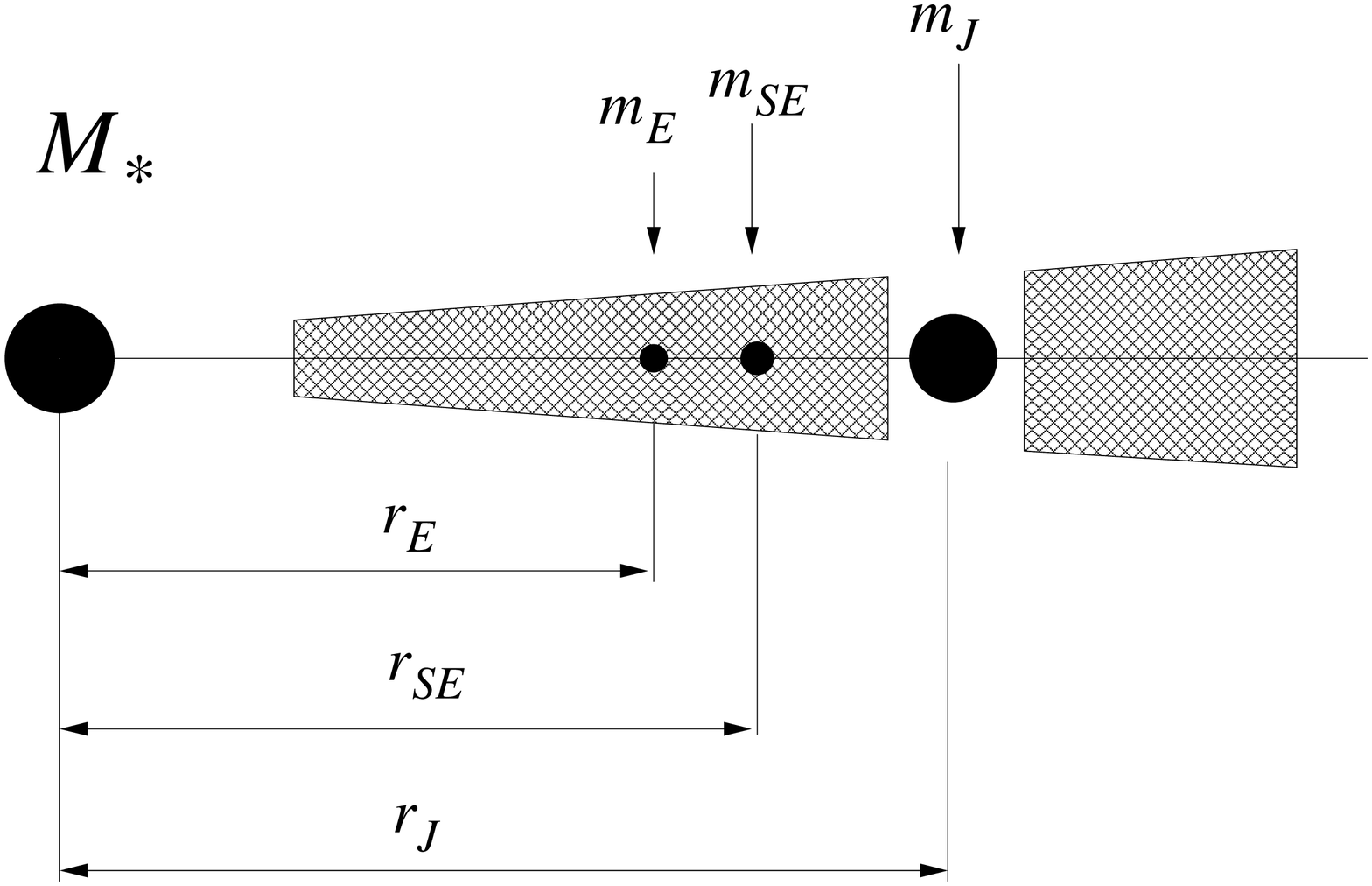}
\caption{\label{fig01}{The initial set up for the numerical simulations
of the two low-mass planets and a gas giant embedded in a protoplanetary disc.
}}
\end{figure}

\begin{figure*}
\begin{minipage}{180mm}
\centering
\vskip 5.5cm
\includegraphics{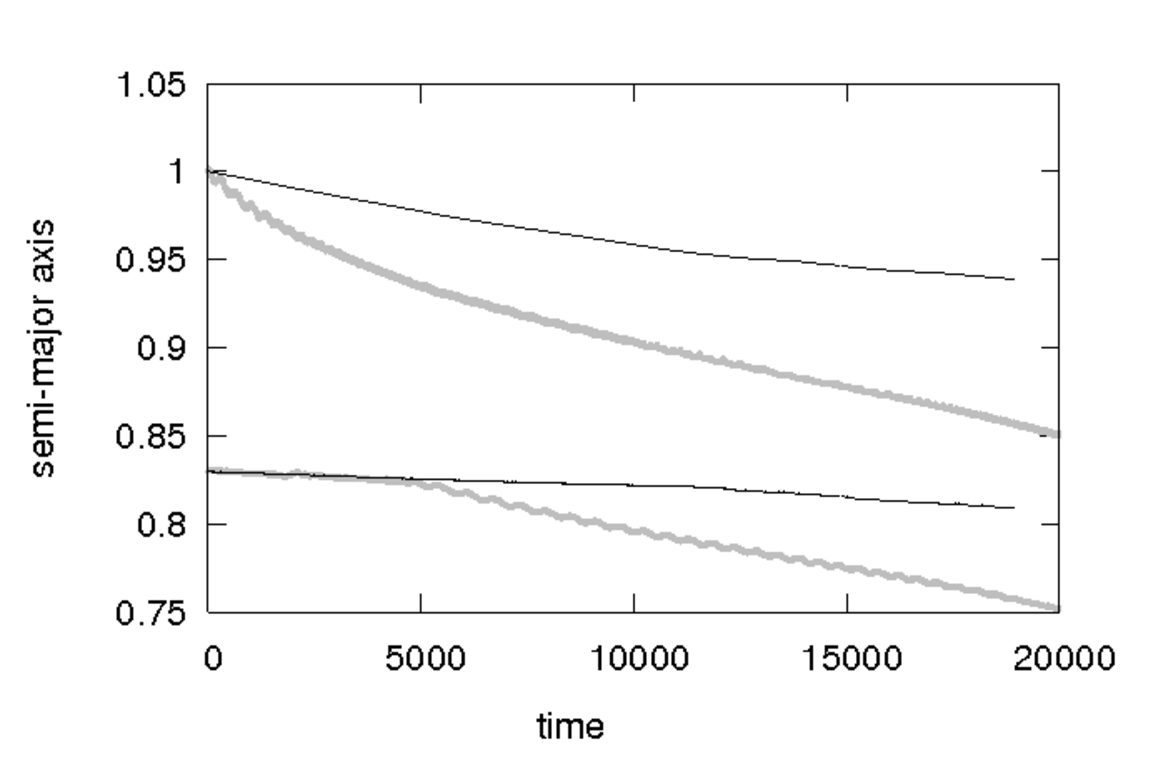}
\includegraphics{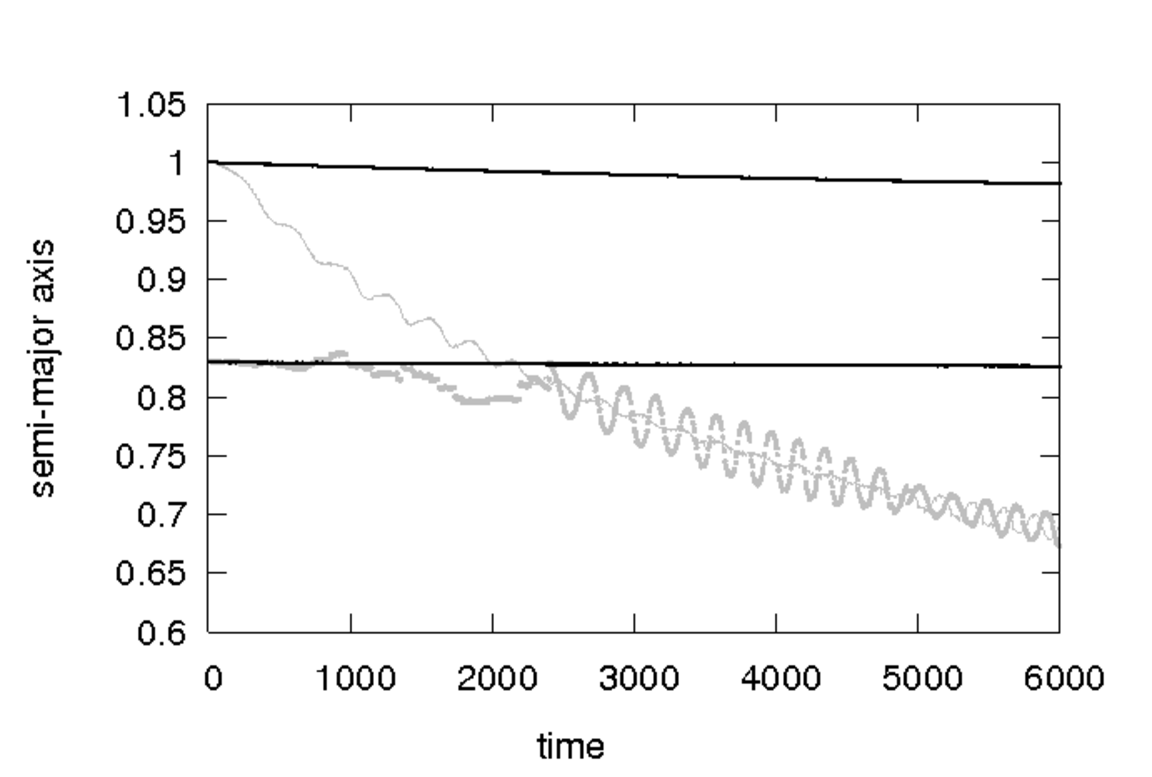}
\end{minipage}
\caption{\label{fig3}{The comparison of the evolution of semi-major axes
  of low-mass planets in the presence of the gas giant
 (gray colour)
 and for the system without the Jupiter (black colour).
The left and right panels show the evolution of low-mass planet
 in the discs with the kinematic viscosities  
$\nu=2 \cdot 10^{-6}$ and $10^{-5}$
 respectively. The location of the Jupiter-like planet is not shown
in these figures. The upper pair of lines refers to the semi-major axis 
of the super-Earth,  
the lower one to that of the Earth-like planet.
}}
\end{figure*}

\begin{figure*}
\begin{minipage}{180mm}
\centering
\vskip 5.5cm
\includegraphics{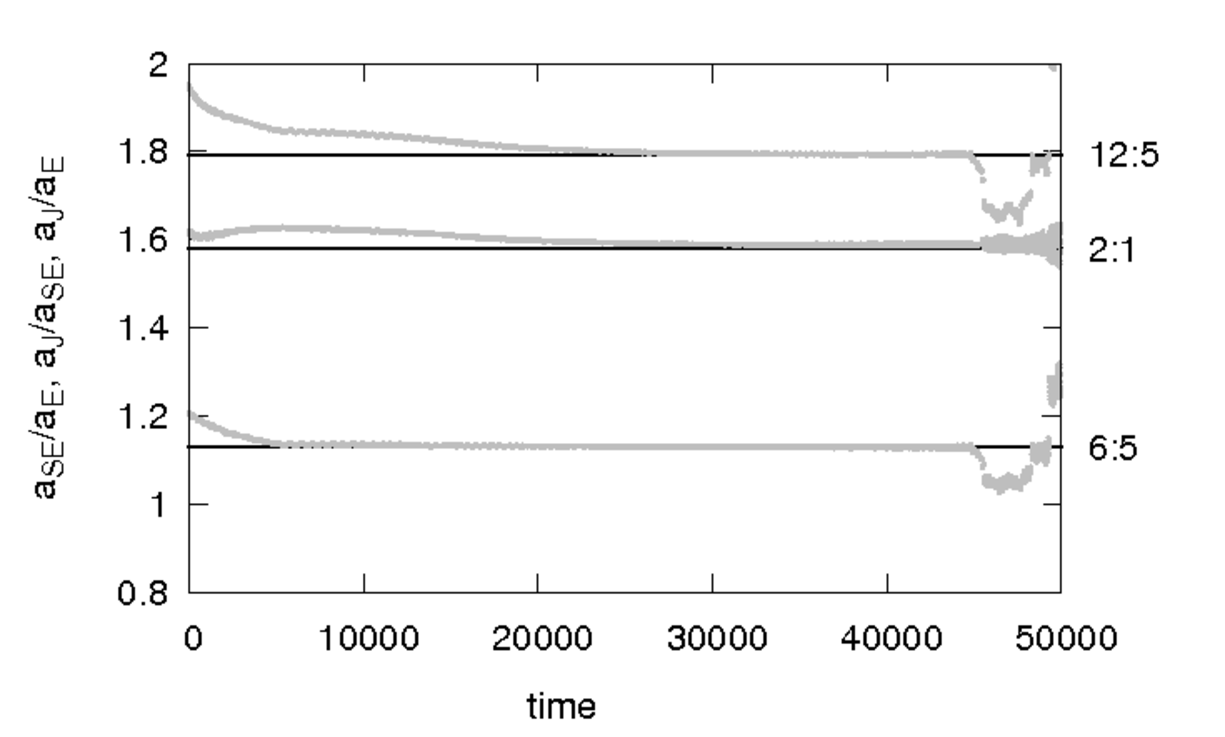}
\includegraphics{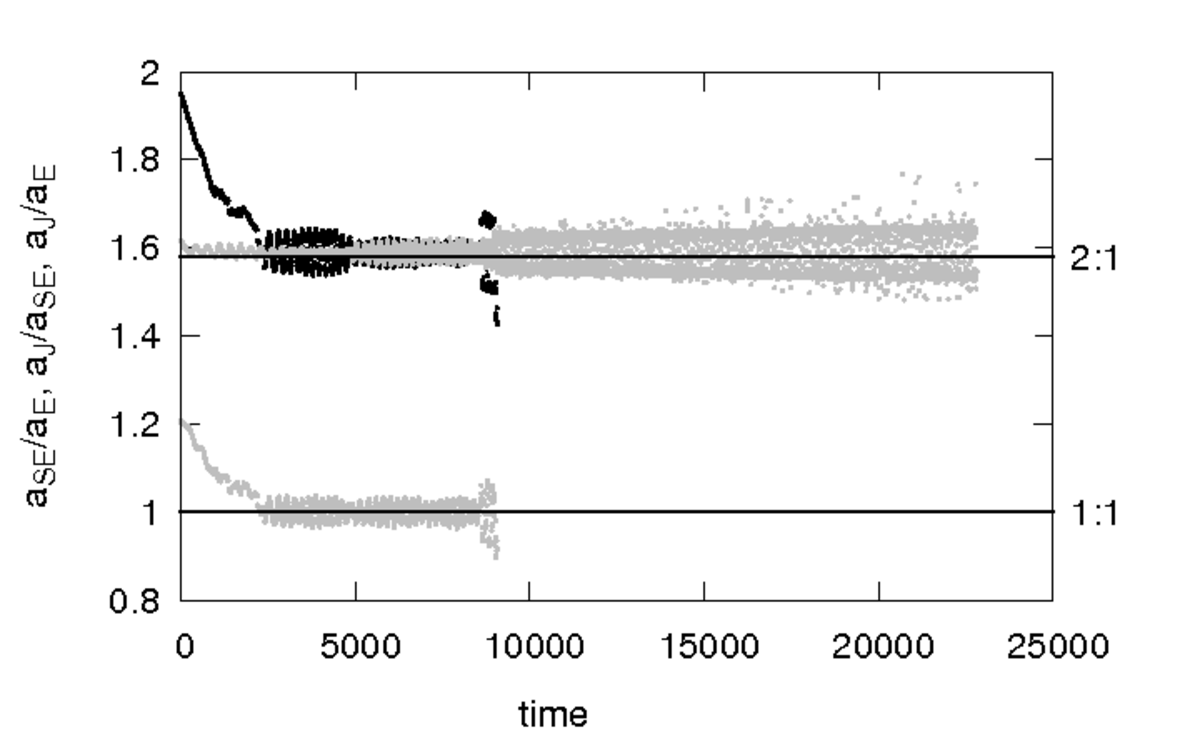}
\end{minipage}
\caption{\label{fig4}{The evolution of the semi-major axis ratios of the
planets embedded in the discs with the kinematic viscosities 
$\nu = 2 \cdot 10^{-6}$
(left panel)
and $\nu=10^{-5}$ (right panel). The horizontal lines denote the exact
position of the resonance.
}}
\end{figure*}

In the previous Section we have presented the most likely resonant 
configuration
for a two low-mass planet system: an Earth analog and a super-Earth
embedded in a locally isothermal gaseous disc with the properties
described in Section \ref{numerics}. Here, we show  
how the presence of a gas giant  can influence the evolution
of such a system. 
The
giant planet is located initially on the external orbit (see Fig. \ref{fig01})
 outside any first
order mean-motion resonance and migrates with  speed 
determined by the viscosity of the disc.
In our simulations we consider two values of  $\nu$, namely $2\cdot
10^{-6}$ and 
$10^{-5}$. Fig. \ref{fig3} shows how the gas giant modifies the
migration rate of the super-Earth. The left panel of that Figure displays how
the semi-major axes of the planets' orbits evolve in a disc with viscosity 
$2\cdot 10^{-6}$. The same information, but
for a disc with $\nu=10^{-5}$, is presented in the right panel. The initial
evolution of the semi-major axes of the two low-mass planets in the
presence of 
the gas giant (gray colour) is compared with the situation in which
there is no massive 
planet in the system (black colour). 
The presence of the gas giant in the system makes the migration of 
the $4 M_{\oplus}$ 
planet  faster than before. This results in turn
in a faster differential migration of the two low-mass companions. 
The relatively slow migration of the gas giant, which occurs in the 
disc with 
$\nu=2\cdot 10^{-6}$, 
causes the low-mass planets to pass through the 5:4 mean-motion
commensurability without
being captured in it. They  proceed instead to another first order 
commensurability, 
namely the 6:5.  
In the disc with higher viscosity ($\nu=10^{-5}$),  the Jupiter migrates
very fast and this leads to the 1:1 resonant locking between the super-Earth 
and the Earth analog.
In the left and right
panel of Fig. \ref{fig4} we plot
the evolution of the semi-major axis ratios of the planets in the 
discs with the
kinematic viscosities $2\cdot 10^{-6}$ and $\nu=10^{-5}$ respectively. 
The aim of these plots is to show what will be the final outcome of the
evolution of the two systems described above: Will the 6:5 resonance
actually take place and
will the 1:1 resonance configuration survive during further migration?  

\begin{figure*}
\begin{minipage}{180mm}
\centering
\vskip 5.5cm
\includegraphics{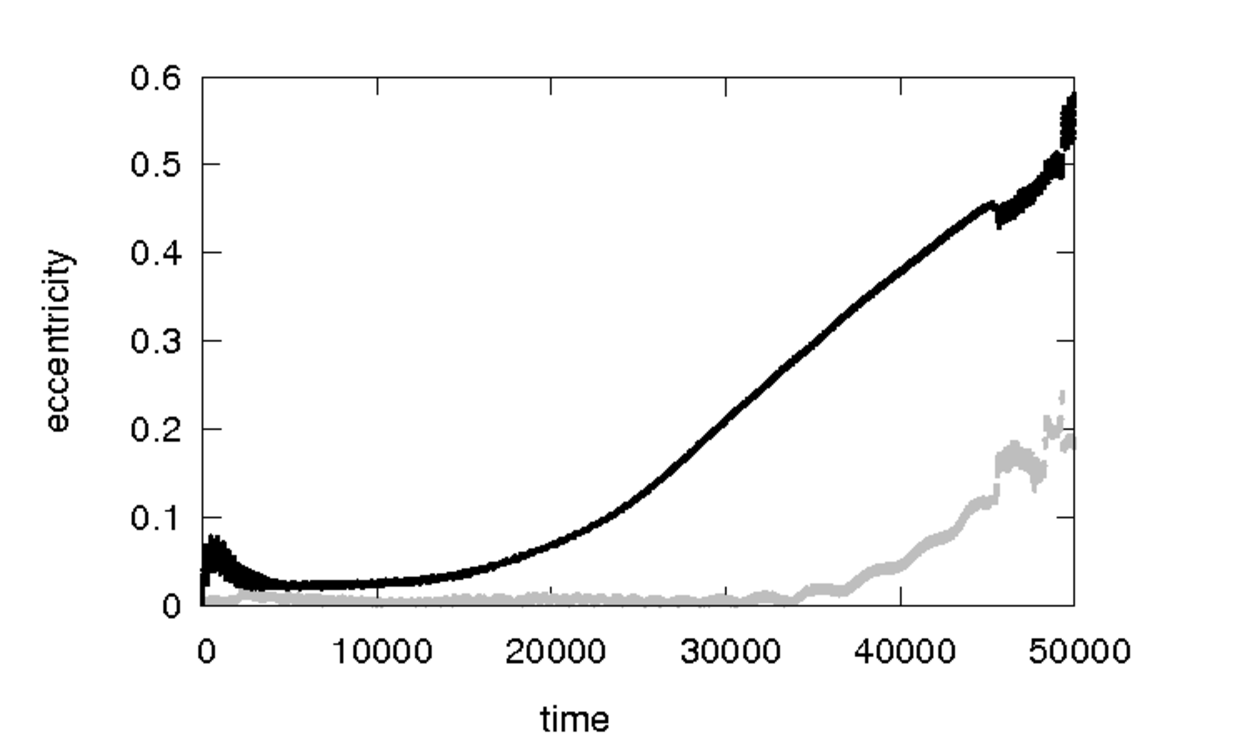}
\includegraphics{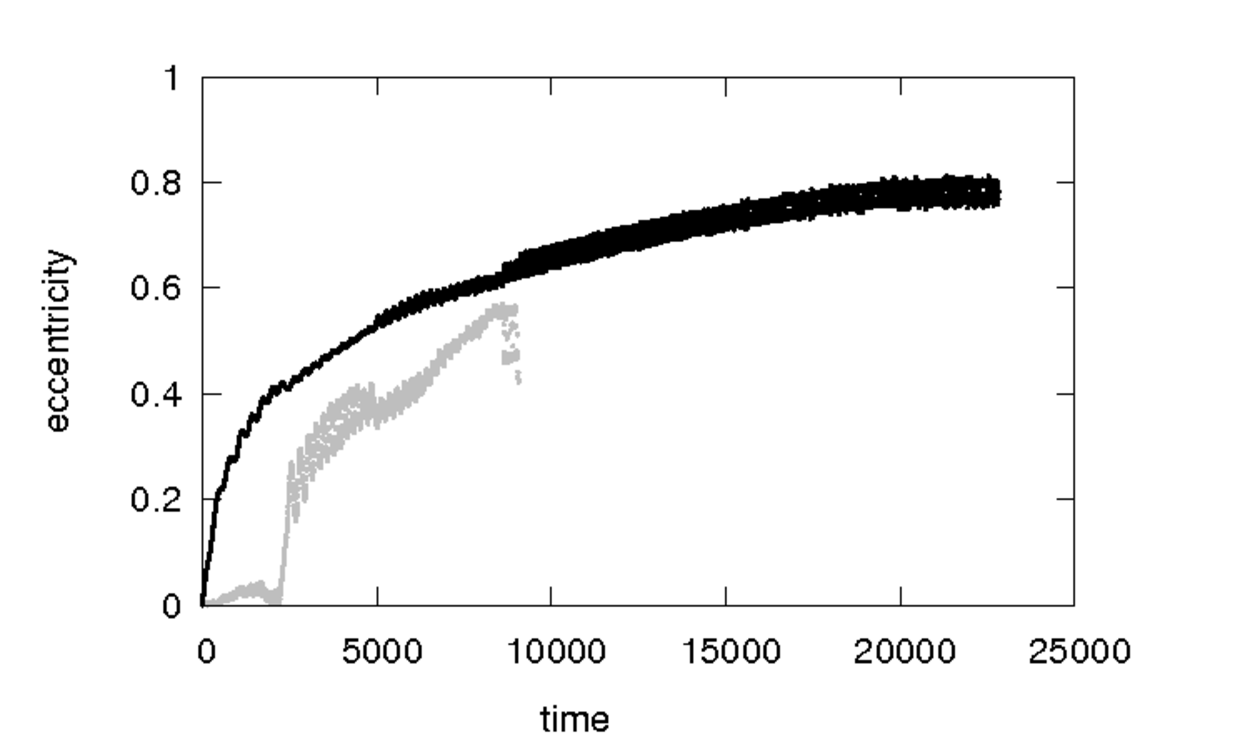}
\end{minipage}
\caption{\label{fige}{The evolution of the eccentricities of the
Earth-like planet (gray curve)
and the super-Earth (black curve) in the disc with the viscosities
$2\cdot 10^{-6}$ ({\it left panel})
and $10^{-5}$ ({\it right panel}).
}}
\end{figure*}

\begin{figure}
\centering
\vskip 5.5cm
\includegraphics{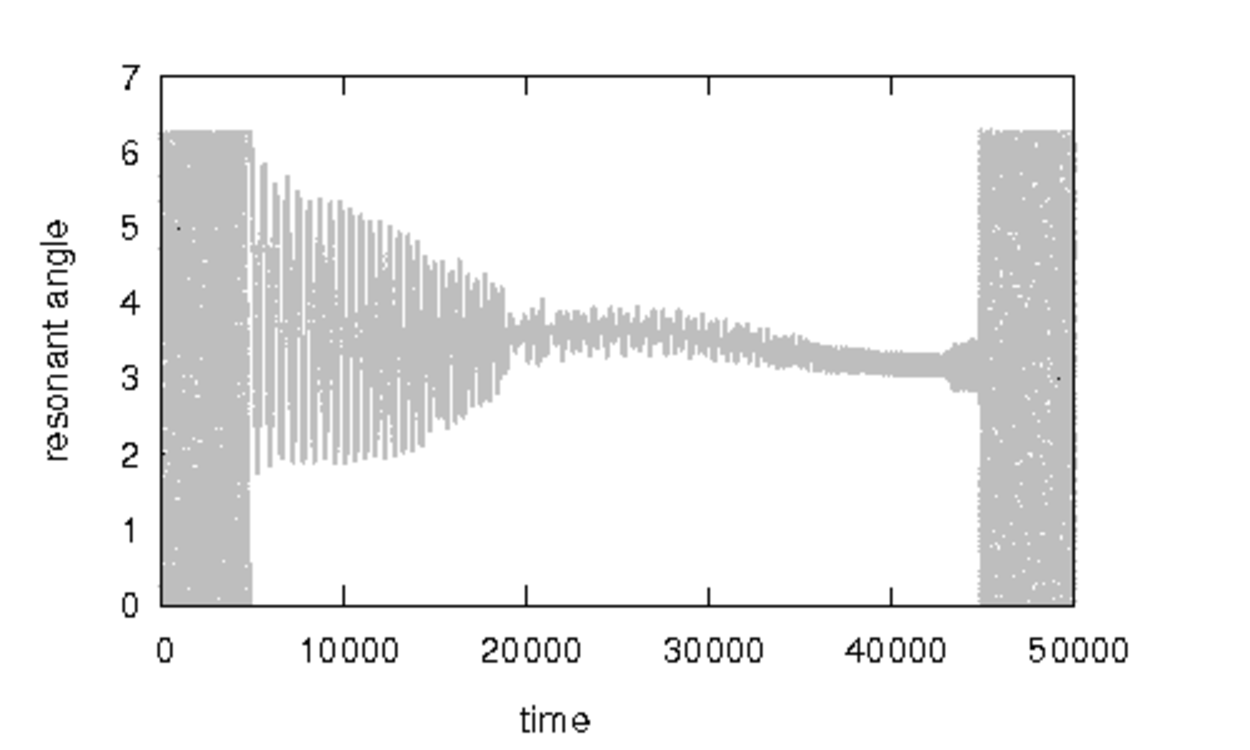}
\caption{\label{fig65}{The evolution of the 6:5 resonance angle 
for the Earth-like planet
and the super-Earth embedded in the disc with the viscosity $2\cdot 10^{-6}$,
migrating in the vicinity of the Jupiter-like planet.
}}
\end{figure}

\begin{figure*}
\begin{minipage}{180mm}
\centering
\vskip 20.0cm
\includegraphics{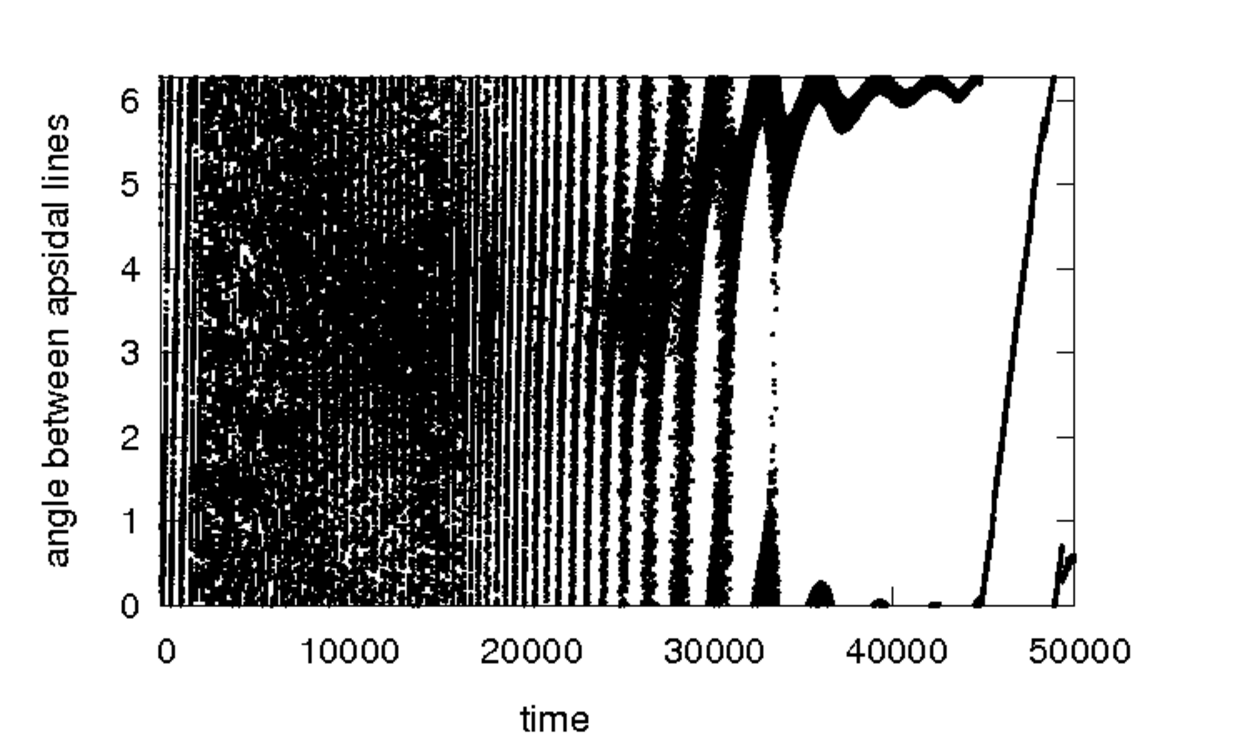}
\includegraphics{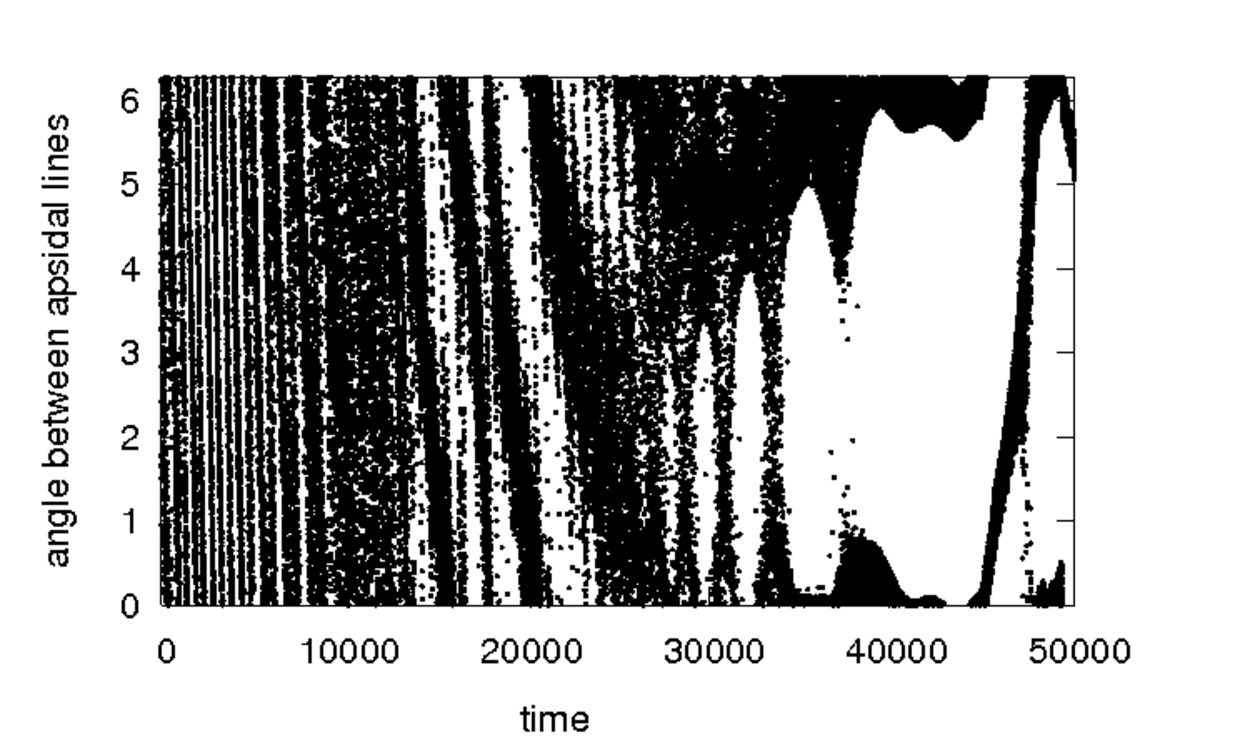}
\includegraphics{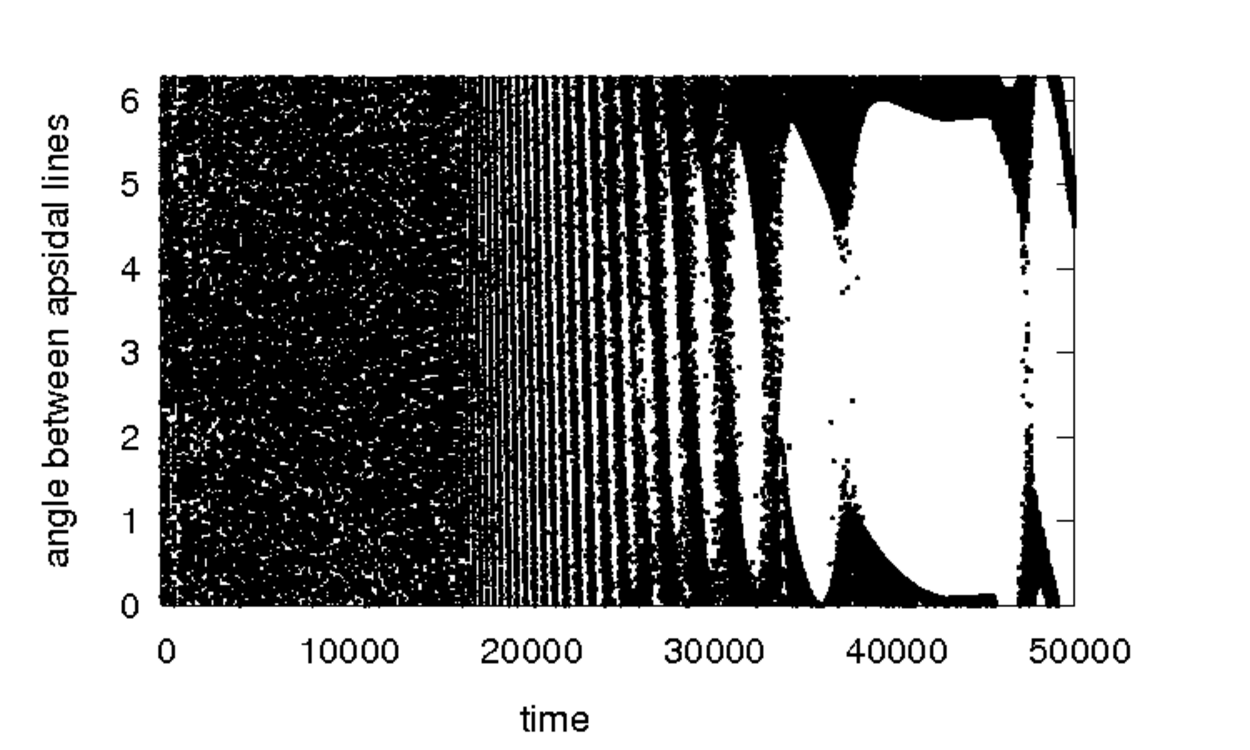}
\includegraphics{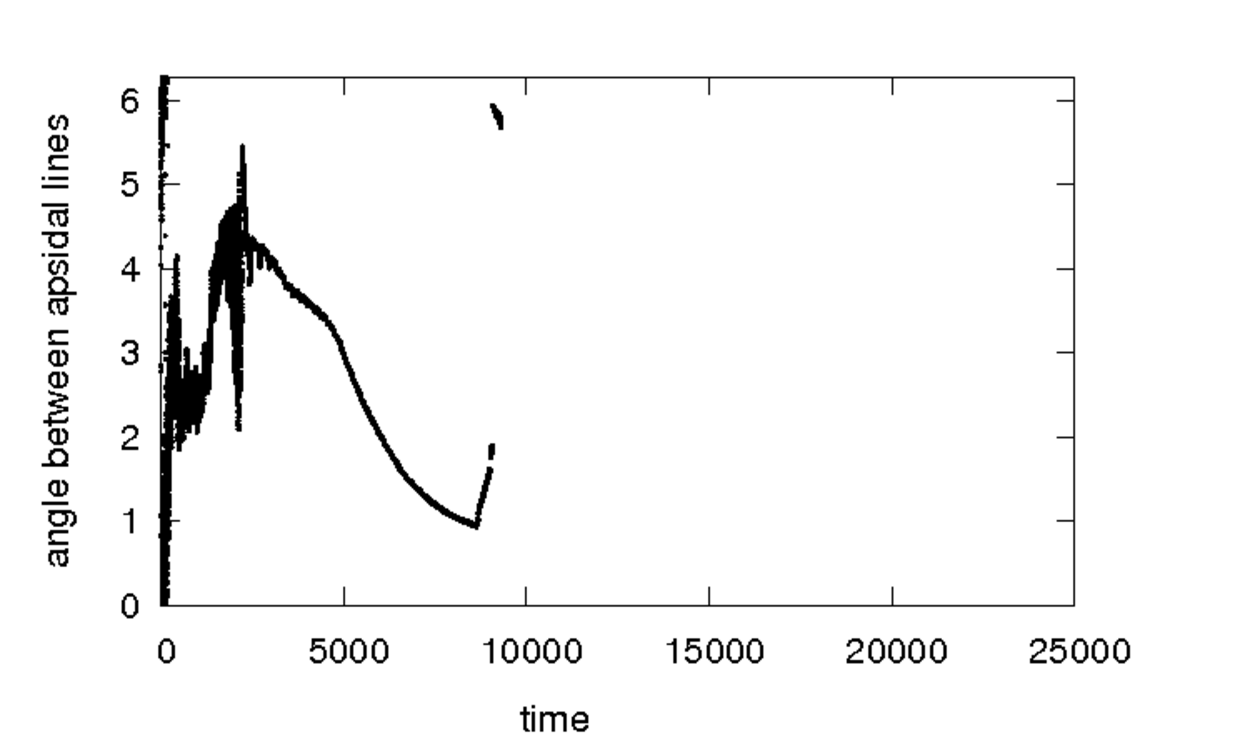}
\includegraphics{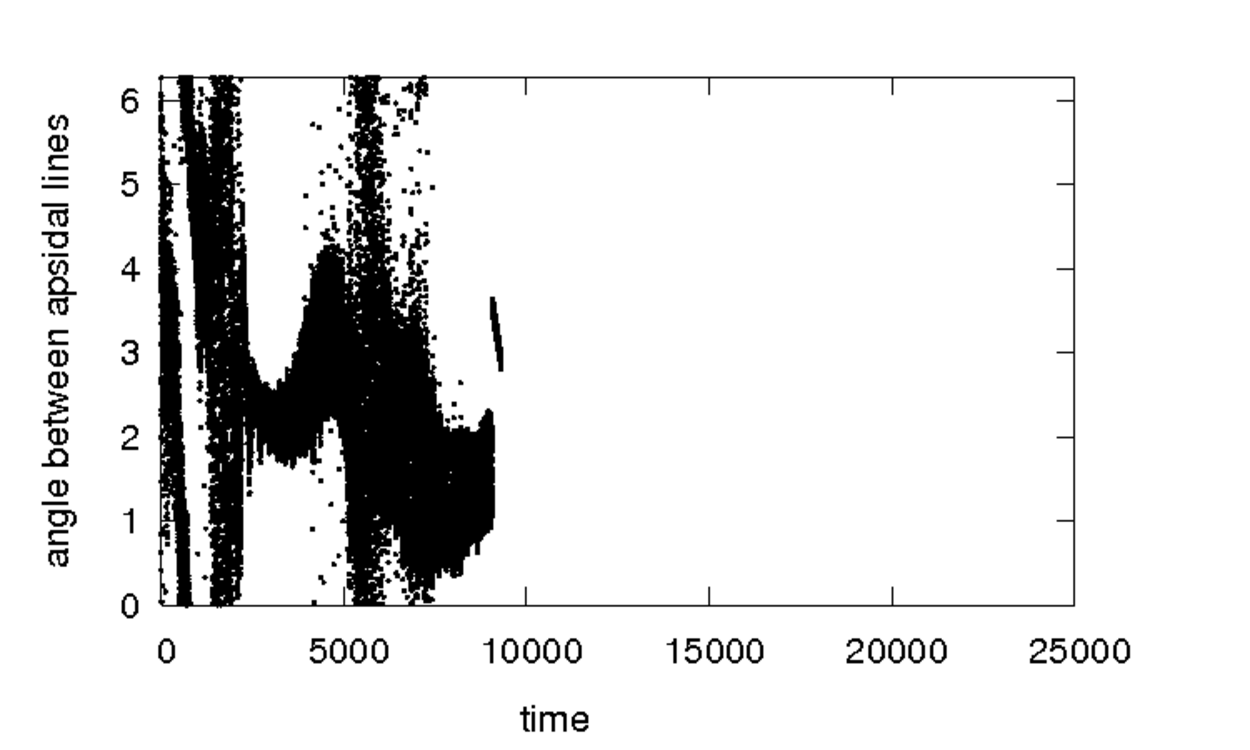}
\includegraphics{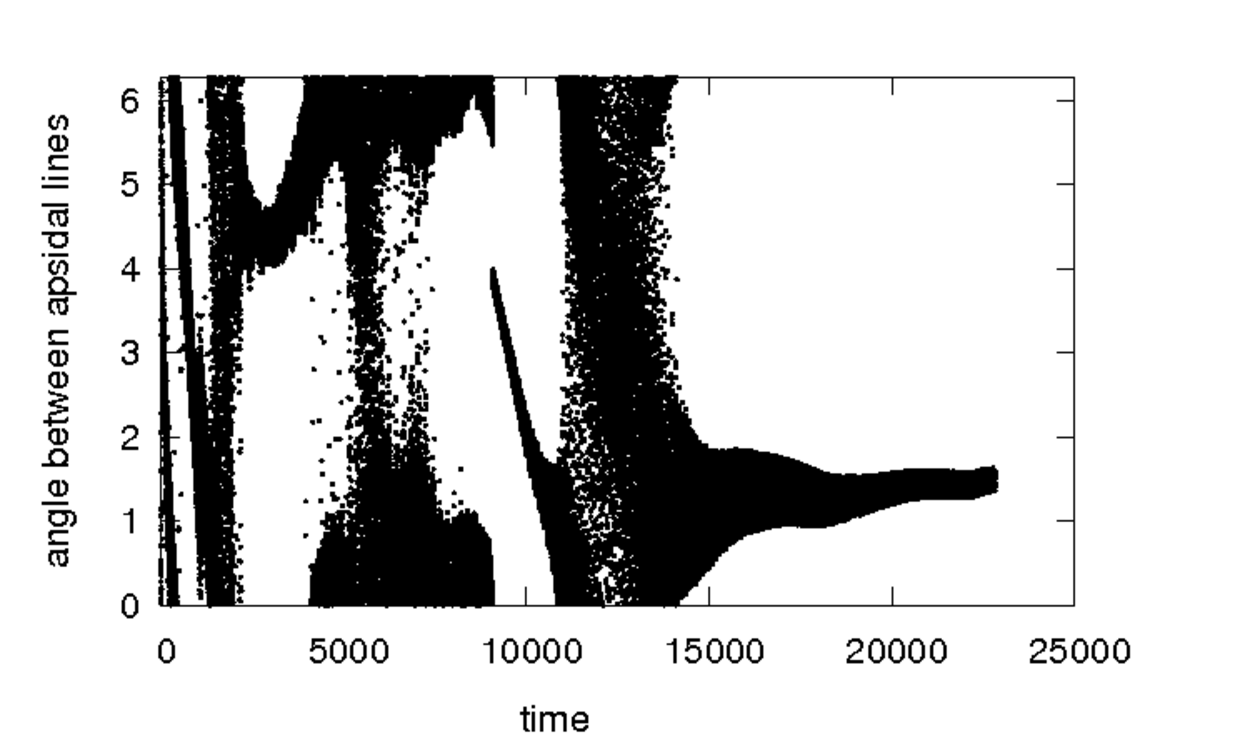}
\end{minipage}
\caption{\label{figaps1}{{\it Left panel}: 
The evolution of the angles between apsidal lines of the Earth-like planet
and the super-Earth ({\it top panel}), the Earth-like planet and the gas 
giant ({\it middle panel}),
and the super-Earth and the gas giant ({\it bottom panel}) in the disc 
with the viscosity $2\cdot 10^{-6}$. 
{\it Right panel}: The same as in left panel but 
for the disc with the viscosity $10^{-5}$. 
}}
\end{figure*}

Let us start from the case in which the two low-mass planets are approaching 
the 6:5 resonance, i.~e. when they  are
embedded in a disc with viscosity
$\nu=2\cdot 10^{-6}$. At the beginning of the simulations, the gas giant is 
placed very close to the 2:1 resonance with the super-Earth. For the first
several hundred time units of its evolution, the differential migration 
of these two
planets is convergent. There is a sign of the excitation of the super-Earth eccentricity,
which is clearly seen in Fig. \ref{fige} (left panel) and which is due to the
vicinity of the 2:1 resonance. However,  the capture into the 2:1 commensurability
did not take place at this early stage, because, after the short period of the convergent
migration, the planets have migrated divergently for several hundreds of time
units. Only
after that time the process of the formation of the 2:1 resonance started and
we have to wait for its completion  for
almost 40000 time units. 
The low-mass planets have reached a position
close to the resonance after approximately 6000 time units. This is
revealed by the libration of the resonance angle
$\phi =6 \lambda_{SE} - 5\lambda_E -\omega_E$ around a fixed value.
Here $\lambda_{SE}$ and $\lambda_E$ denote the mean longitudes of the  
super-Earth
($SE$) and the Earth analog ($E$) respectively, while 
$\omega_E$ is the 
longitude of the pericentre of the Earth analog.
Initially the libration amplitude is large (170 degrees), but by the time
of 20000 time units it decreases to 50 degrees and after  35000 time units 
is around 25 degrees, as it can be seen from Fig. \ref{fig65}. The appearance of the
resonant angle defined above is not associated with a significant
increase in the eccentricities of the planets' orbits.
The signature of the 6:5 resonance is present also in
the behaviour of the angle between the apsidal lines, shown 
in  Fig. \ref{figaps1} (left panel). 
In this way, the system has attained 
the configuration in which the Jupiter and the super-Earth
are  locked in the 2:1 commensurability and, in addition, the super-Earth and
the Earth analog are locked in the 6:5 commensurability.
This means that the resonant configuration is established also between
the Jupiter and the Earth-like planet. Their period ratio equals 12:5.
Thus, the three planets in the system are locked in the triple 12:6:5 
mean-motion resonance. The illustration of this statement is presented in 
Fig. \ref{figaps1}.
However,
such resonant configuration is not  maintained during the further evolution. 
Soon after the capture in the triple resonance, the orbits of both low-mass 
planets become chaotic (see Fig. \ref{fig4}, left panel, the uppermost and 
lowest lines
almost at the end of the displayed period of the evolution)  
and the super-Earth is shifted on the most external orbit. 

\begin{figure*}
\begin{minipage}{180mm}
\centering
\vskip 5.5cm
\includegraphics{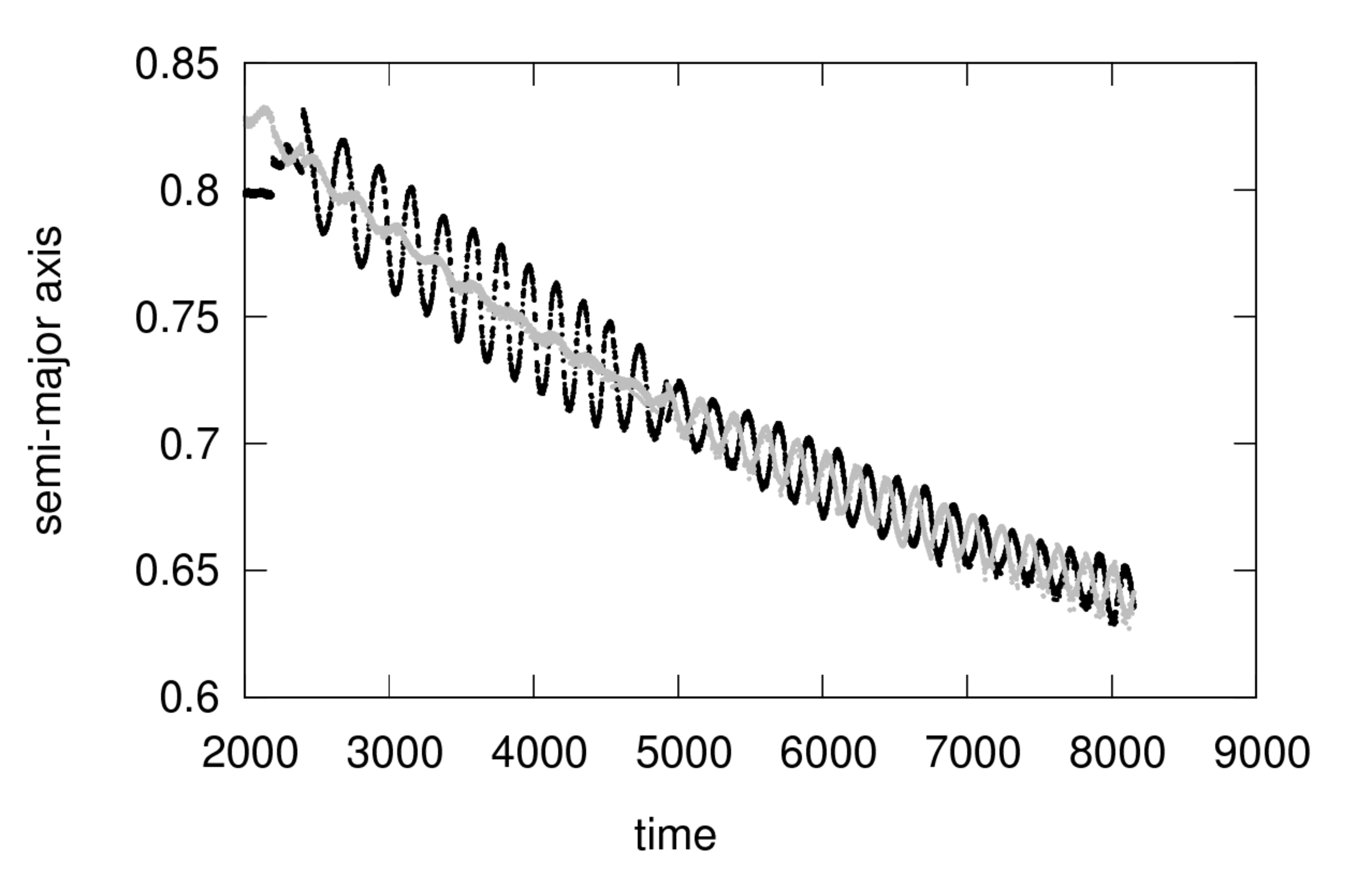}
\includegraphics{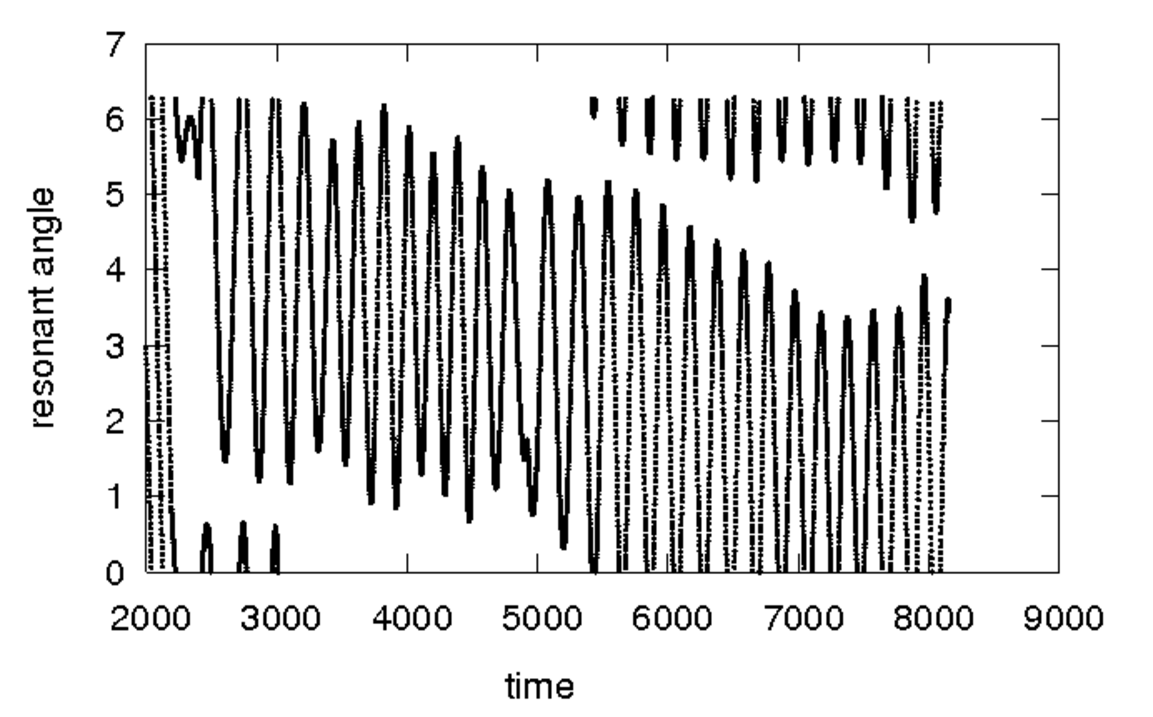}
\end{minipage}
\caption{\label{res11}{The evolution of the semi-major axes of the
planets in the co-orbital motion
({\it left panel})
and the behaviour of the resonance angle ({\it right panel}). 
In the left panel, the gray dots refers to the semi-major axis of 
the super-Earth and
the black ones to that of the Earth-like planet. 
}}
\end{figure*}

On the other hand, in the disc with viscosity $10^{-5}$ the gas giant  
first captures the super-Earth in
the 2:1 commensurability and then the super-Earth 
captures  the 
Earth-like planet
into the 1:1 commensurability. 
As a consequence, the three planets  form
the triple 2:1:1 resonant configuration which, however, does not survive
during the further evolution. 
The lightest planet is ejected from the system after roughly 9000  
time units, see Fig. \ref{fig4}, right panel, lower line.
Despite the scattering occurring in the system, the gas giant and the 
super-Earth 
remain locked 
in the commensurability till the end of the simulation (see Fig. \ref{figaps1}).
The capture in the 1:1 resonance has already been observed in a variety of
numerical studies, see e.g. \cite{thommes},
\cite{beauge}, \cite{crida2009b}, \cite{cresswell2008,cresswell}, \cite{paperIII}. 
Here we
would like to take a closer look at the type of motion of the two low-mass planets
in the co-orbital configuration induced by the vicinity of a migrating gas giant
and affected by interaction of the 2:1 resonance. 
In Fig. \ref{res11} we present the evolution
of the orbits of the two co-orbital planets and the resonance angle, which for the 1:1 commensurability
is just equal to the difference between the mean longitudes of the planets. 
The amplitude of libration of the resonance angle is very large, 
which makes difficult to
determine the type of the planet motion. During the first part of the 
co-orbital evolution, after 
a very short transient event (close encounter), which took place at about
2300 time units,
the planets gradually settled down on 
mutual horseshoe orbits. The resonant angle librates roughly around 180
degrees and the  
relative 
variation of the semi-major axis of the Earth-like 
planet is large. 
At around 5000 time 
units the situation has changed. The amplitude of the semi-major axis 
of the Earth-like planet has decreased significantly and a value
around which the resonant angle librates decreases slowly in time,
reaching towards the end of our simulations 
(between 7000 and 7800
time units) the value about 60 degrees. This could be an indication that
at this stage of the evolution the planets move on the tadpole orbits.
Finally, at around 9000 time units, the
Earth-like planet is ejected from the system, falling into the star. 

The fact that  triple resonances are not maintained might be a
consequence of the eccentricity excitation of the low-mass planet orbits. 
Indeed, after
the capture in the 2:1 resonance, the eccentricities of both low-mass planets 
increase significantly.
For the disc with viscosity $2\cdot 10^{-6}$, the eccentricity of the 
Earth-like planet reaches 0.12 and that of the super-Earth approaches  0.45.
 In the case of the disc with higher viscosity 
($\nu=10^{-5}$), the eccentricities increase to even higher values,
 namely 0.55 and  
0.8
for the Earth analog and the super-Earth respectively. 
The motion on such highly eccentric orbits may lead to the chaotic behaviour
in the system which is
observed in our simulations and may be related to the 
disappearance of the triple resonances.

Thus, we argue that, under the mutual conditions discussed in this Section, 
in systems containing a gas giant and two low-mass
planetary companions, the capture in the triple resonance due to convergent 
migration
can be very common at the early stages of the evolution. However, the final
configuration of such systems consists of a gas giant and  only one 
low-mass planet on the
internal orbit. The second low-mass planet is moved from the internal 
orbit onto the external one or is entirely ejected from the system due to
the eccentricity excitation of the planetary orbits.

%
%
%
%
%

\section{Discussion and conclusions}
\label{conclusions}

In Section \ref{gasgiant} we have considered  
the situation in which  the gas giant is located 
initially outside any first order mean-motion resonance, namely 
at the distance $r_J=1.62$.
Now, we discuss the possible final architectures for a few other initial 
locations of the Jupiter-like planet, assuming that such configurations 
were formed in the previous stages of the evolution of the planetary 
systems.

First, we place the gas giant at the distance $r_J=1.35$.
If the viscosity of the disc is low ($2 \cdot 10^{-6}$), the most inner
planet moves on a chaotic orbit since the beginning of the simulation.
Meanwhile, the Jupiter and the super-Earth approach  the  
3:2 commensurability. 
The eccentricity of the super-Earth increases only to about 0.1, which is lower 
than what was observed for this resonance in our previous simulations 
\citep{paperI}. Moreover, the angle
between the apsidal lines oscillates in the whole range from $0$ to $2\pi$. 
Finally, 
after 16000 time units both low-mass planets are scattered from the disc.
If instead the viscosity of the disc is high ($\nu=10^{-5}$), the convergent
 migration leads to the temporary capture in the triple 4:3:2
mean-motion resonance. During the evolution of such configuration, 
the eccentricities of the low-mass  planets increase and, as a result,
the Earth analog is ejected from the system. This event has a big effect 
on the system, because 
the commensurability between the Jupiter and the super-Earth is destroyed.
Soon after this happens, also the second low-mass planet is scattered 
from the disc.
The final result in both cases, regardless of the value of the 
viscosity, is just a single gas giant orbiting 
the star. 

In the next simulation,
the gas giant is placed initially even closer than before, i.e. 
at $r_J=1.28$. In this case, we find that 
in the disc with low viscosity ($\nu=2 \cdot 10^{-6}$), the Jupiter
captures 
the super-Earth into the 3:2 resonance. Then, the Earth-like planet is ejected 
from the system and  the resonant structure between the gas giant
and the super-Earth is lost. However, during the further evolution of 
the system,
the convergent migration brings the super-Earth and the gas giant back 
into that 
commensurability.

In the disc with viscosity $10^{-5}$, the super-Earth is scattered
into the most external orbit at a distance close to $r=2$ and slowly
migrates towards the Jupiter. We expect that the super-Earth will stop to
migrate just outside the 2:1 external resonance with the gas giant 
as it has been shown in \cite{paperIV}. The  super-Earth still did
not reach the predicted location at the time of the end of simulations. 
The Jupiter, instead, captures the Earth analog 
in the 2:1 resonance and this commensurability survives  till the end of the 
simulation.  
Again, the final configuration consists of  the
gas giant and  only one low-mass planet on the internal orbit.

Concluding, in Section 3 we have investigated systems of two planets with 
masses 1M$_{\oplus}$ and  4M$_{\oplus}$  embedded into a locally
isothermal disc.
It has already been  demonstrated that
for such systems, evolving in the inviscid disc,
the capture into a first
order mean-motion resonance is very likely if the relative migration of the 
planets is 
convergent
\citep{papszusz, papszusz10}. We confirm here this result also for  
discs with 
viscosities in the range between $0$ and $10^{-5}$.
In all our simulations the planets attain finally the
5:4 resonance, even if the path through which it occurs can be very different
from one case to another.

The main goal of this study was to investigate 
how the architecture of a system  changes
if it contains not only a pair of  low-mass planets, but also a Jupiter mass 
gas giant located on the most external orbit. Shall we expect to assist to the
formation of a resonant configuration in such systems?  
Apart from those cases in which both or one low-mass planets are ejected
from the system at the very early stages of the simulations, the most common 
configuration is a triple commensurability.
However, in all our simulations this is just a transitional configuration.
It never survives for a long time.
Soon after the resonant locking, 
the eccentricity of both low-mass
planets increases and  close encounters between planets occur.
Finally,  the 
ejection of one or
both low-mass planets from the internal orbits takes place.
The most likely final scenarios of the
evolution of the systems described in this paper, for given planetary 
masses and initial planet locations, are that
i) only the
giant planet is left in the system or
ii) a resonant pair of planets remains that can consist in one of the
following: 
The gas giant 
and the super-Earth on the
internal orbit or the gas giant
and an Earth analog on the internal orbit. 

Within our simulations we have observed also  cases  
in which a low-mass planet moves from the internal to the external orbit due 
to the
action of the Jupiter-like planet. Interestingly enough, the low-mass planet
is located close to the 2:1 resonance but not exactly in resonance. 
This might be
a consequence of the interaction of the low-mass planet  with the
density waves 
excited by the gas giant described in details in \cite{paperIV}.
To sum up, it is important to stress here that the variety of outcomes
described above is mainly  due to the action of the gas giant. 
Because of the presence of the Jupiter in the disc, in fact,
the  
initial migration rates of the low-mass planets  undergo very
quickly  an alteration. 
After that, the motion of the low-mass planets  
is dominated  by the interaction with the gas giant and the
waves excited by it in the disc. 

Our conclusions are concerning a system of three planets embedded in a
locally isothermal gaseous disc. One needs to be aware of all limitations these
assumptions can bring. 
Let us  present a straightforward argument, which  will help  to 
justify the use of the locally isothermal equation of state in our 
simulations. 
To this purpose,  we have first 
evaluated the opacity of the disc at the initial location of the super-Earth, 
namely at $r=5.2$ AU. Adopting  the values of surface density, aspect ratio, 
and distance 
considered in our paper, we have found that  the disc is optically thick 
at that position. 
Next, we have evaluated 
the migration rate from  equations (50-53) in   \cite{Paa2010b},
which give the total torque acting on a low-mass planet consisting of the
Lindblad torque plus the corotation torque. Finally, we have compared this
migration rate with that obtained in the case of the locally isothermal limit 
using equation (49) in   \cite{Paa2010a}, 
getting an agreement between the two values within a few percents.
This upshot confirms that we
have studied here a particular case of the migration rates of the planets
where the corotation torque can be neglected, namely the planets, which
move in  an optically thick
disc  well modelled by the locally isothermal approximation. 
Other disc 
conditions may lead to other migration rates (in particular for the 
terrestrial planets), see \cite{Bitsch13}.  

The most obvious extension of this work will
be to move into a more realistic equation of state or ideally to use a self-consistent
radiative model of the protoplanetary discs. This will allow us to understand how the 
low-mass planets migrate in the  given physical conditions occurring in the
observed discs. Much work has already been done towards this aim (see for
instance \cite{PaarMel2006, PaarPap2008, KleyCrida2008, BarMas2008, Bitsch10, 
Bitsch11a,
Bitsch11b}. 

So far, we know  examples of systems in which low-mass planets are 
close to the 5:4
commensurability (Kepler-11, \cite{lissauera}) or form triple 
resonances or even
multiple
commensurabilities involving a 1:1 resonance \citep{lissauerb, laughlin}. All 
these structures are seen also in our numerical simulations at some stage
of the evolution. 
On the other side,  
neither a super-Earth nor an Earth analog orbiting
on the internal orbit relative 
to a gas 
giant and 
locked with it in a mean-motion resonance have been detected yet. 
However, our simulations   
indicate that 
these
configurations are quite likely to form
 at the early stages of the planetary system evolution.
The increasing number of data
(from the numerous planet searches) should shed more light on the existence
of such configurations soon.

\section*{Acknowledgments}
This work has been partially supported by the MNiSW grant N203 026 32/3831
and  the MNiSW PMN grant - ASTROSIM-PL ''Computational Astrophysics.
The formation  and evolution of structures in the universe: from planets to
galaxies''.
The simulations reported here were performed using the HPC cluster "HAL9000"
of the Faculty of Mathematics and Physics at the University of Szczecin.
We are grateful to Micha{\l} R\'o\.zyczka for bringing this topic to our 
attention and	
Franco Ferrari for reading the manuscript and 
his continuous support in the development of our
computational 
techniques and computer facilities.
Finally, we thank the referee for valuable comments which helped
us in improving the manuscript.

\label{lastpage}

\end{document}